\documentclass{svproc}
\usepackage{url}
\usepackage{graphicx}

\begin{document}

\mainmatter              

\title{Anisotropic flow predictions for identified and strange hadrons in $O+O$ collisions at $\sqrt{s_{\mathrm{NN}}}$ = 7 TeV using model approaches}

\titlerunning{Anisotropic flow in $O+O$ collisions}

\author{J. Singh\inst{1} \and M. U. Ashraf\inst{2} \and A. M. Khan\inst{3} \and S. Kabana\inst{1}}
\tocauthor{M. U. Ashraf, A. M. Khan and S. Kabana}
\institute{
  Instituto de Alta Investigaci\'on, Universidad de Tarapac\'a, Casilla 7D, Arica, 1000000, Chile\\
  \and
  Department of Physics and Astronomy, Wayne State University, 666 W. Hancock, Detroit, Michigan 48201, USA \\
  \and
  Georgia State University, Atlanta, GA 30303, USA \\
  \vspace{0.3cm}
  \email{Email: jsingh2@bnl.gov}    }

\maketitle     

\begin{abstract}
  In this study, we report the predictions for the flow observables for different centrality classes in $O+O$ collisions. Our predictions utilize two different approaches, hydrodynamic and transport models, to analyze the behavior of the flow coefficients for identified ($\pi^\pm$, $K^\pm$ and $p (\overline{p})$) and strange ($\mathrm{K}^{0}_{\mathrm S}$, $\Lambda$ ($\overline{\Lambda}$), $\Omega^{-}$ ($\overline{\Omega}^{+}$), $\Xi^{-}$ ($\overline{\Xi}^{+}$), $\phi$) hadrons. We explore particle-by-particle flow and compare the response of the system to initial conditions across various models, which provide insights into the underlying partonic and hadronic dynamics. The study presents comparisons of flow harmonics with the existing experimental measurements and demonstrates how $O+O$ collisions can serve as a benchmark to understand the transition from small to large systems, contributing to our knowledge of the Quark-Gluon Plasma (QGP) and collective phenomena in heavy-ion collisions.
\end{abstract}

\section{Introduction}
The primary objective of the heavy-ion physics programs at the Relativistic Heavy Ion Collider (RHIC) and the Large Hadron Collider (LHC) is to explore the characteristics of nuclear matter under extreme conditions, specifically focusing on the quark–gluon plasma (QGP). The QGP is a state of matter formed at extremely high temperatures and energy densities, where quarks and gluons are no longer confined within hadrons and exist freely~\cite{QGP1,QGP2}. Anisotropic flow studies in heavy-ion collisions provide critical insights into the transport properties of the QGP and the role of initial-state fluctuations. This phenomenon arises from the pressure gradients generated by the asymmetric spatial configuration of the colliding nuclei in the initial stage of the collision. Anisotropic flow is commonly described using flow harmonics, $v_n$, which are the Fourier coefficients of the azimuthal particle distribution and are expressed as $v_{n} = \langle \cos [n(\varphi - \Psi_{n})] \rangle$. Here, $n$ represents the order of the flow harmonic, $\varphi$ denotes the azimuthal angle, and $\Psi_{n}$ is the symmetry plane angle associated with harmonic $n$. The first four Fourier coefficients, $v_1$, $v_2$, $v_3$, and $v_4$, correspond to directed, elliptic, triangular, and quadrangular flow, respectively. The observed anisotropic flow provides strong evidence of collective behavior in the system~\cite{Ollitrault} and has been extensively investigated at the RHIC~\cite{STAR:2005,PHENIX:2004} and the LHC~\cite{ALICE:2010}.

The LHC is expected to collect data from $O+O$ collisions, offering immense potential to deepen our understanding of these phenomena. The multiplicities in $O+O$ collisions are comparable to those in peripheral $Pb+Pb$ and $Xe+Xe$ collisions, providing a unique opportunity for a systematic exploration of initial collision geometries across different nuclear systems. The significant overlap in multiplicity between $O+O$ collisions and with small and large systems presents a compelling opportunity to investigate the relative contributions of initial and final-state effects. This study investigates flow harmonics $v_n$ as a function of multiplicity and $p_T$ in $O+O$ collisions at $\sqrt{s_{\mathrm {NN}}}$ = 7 TeV using recently updated EPOS4~\cite{epos4_1} framework and AMPT~\cite{ampt_1} model, which effectively describes the anisotropic flow across different system sizes i.e., for small systems as well as large systems.

\section{Event generators}
A Multi-Phase Transport (AMPT)~\cite{ampt_1} is a well-established model commonly used to investigate the dynamics of relativistic heavy-ion collisions. In this work, both the default (AMPT-Def) and the string melting (AMPT-SM) versions have been employed to produce simulations. EPOS4~\cite{epos4_1} is a multipurpose event generator which simulates $p+p$, $p+A$, and $\mathrm{A+A}$ collisions using a 3+1D viscous hydrodynamic approach~\cite{epos4_2}. In this study, we employed the Q-cumulant method~\cite{Qcumulant} to efficiently compute multi-particle correlations while minimizing statistical uncertainties and non-flow effects. This technique enhances the accuracy of anisotropic flow harmonic ($v_n$) measurements while optimizing computational efficiency.
A total of $\sim$4 million minimum-bias events were simulated from AMPT-SM and AMPT-Def while $\sim$1.5 million  minimum-bias were simulated from EPOS4.

\section{Result and Discussions}
\begin{figure}[!]
  \vspace{-.02cm}
  \centering
  \includegraphics[width=.65\textwidth]{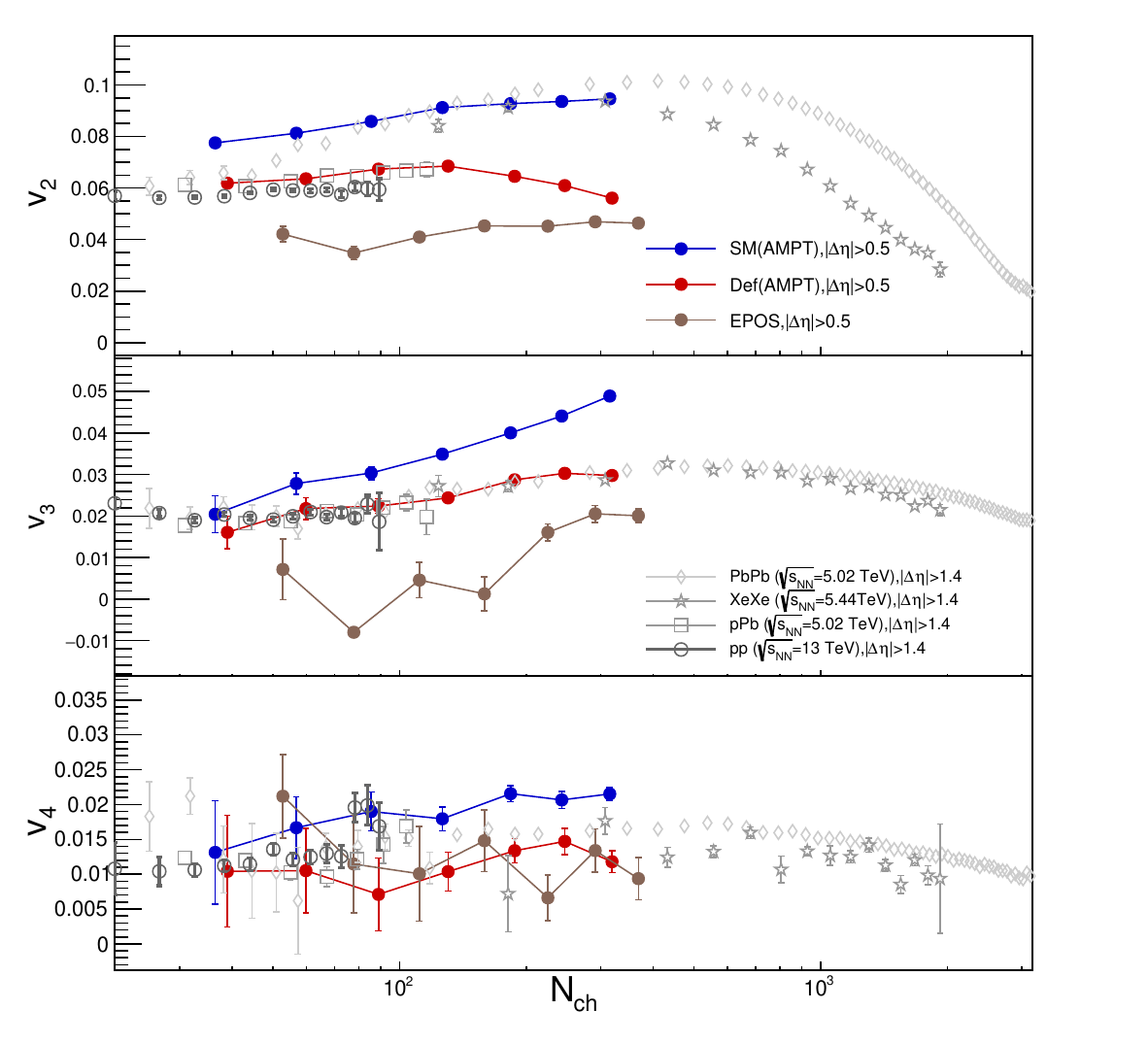}
  \caption{(Color online) Anisotropic flow coefficients ($v_2$, $v_3$, and $v_4$) as functions of charged particle multiplicity ($N_{ch}$) in $O+O$ collisions at $\sqrt{s_{\mathrm{NN}}} = 7$ TeV, shown for the EPOS4, AMPT-Def, and AMPT-SM models.}
  \vspace{-.02cm}
  \label{fig:v2v3v4}
\end{figure}

Figure~\ref{fig:v2v3v4} shows the predictions for $v_2$, $v_3$, and $v_4$ as a function of charge particle multiplicity ($\langle N_{ch} \rangle$) in $O+O$ collisions at $\sqrt{s_{\mathrm{NN}}} = 7$ TeV using EPOS4 and AMPT. These predictions are compared with experimental data for small ($p+p$ and $p+Pb$) and large ($Xe+Xe$ and $Pb+Pb$) systems at available energies.
\vspace{-.02cm}
Our results for $O+O$ collisions show a clear final state multiplicity overlap with exisiting $p+p$, $p+Pb$, and $Pb+Pb$ experimental data. This observed overlap indicates strong final-state interactions and collective behavior, which may hint at the possible formation of a strongly interacting medium with QGP-like properties. Detailed multiplicity tables for all models can be found in Ref.~\cite{OO7_identified,OO7_strange}.

\begin{figure}[!]
  \vspace{-.02cm}
  \centering
  \includegraphics[width=.49\textwidth]{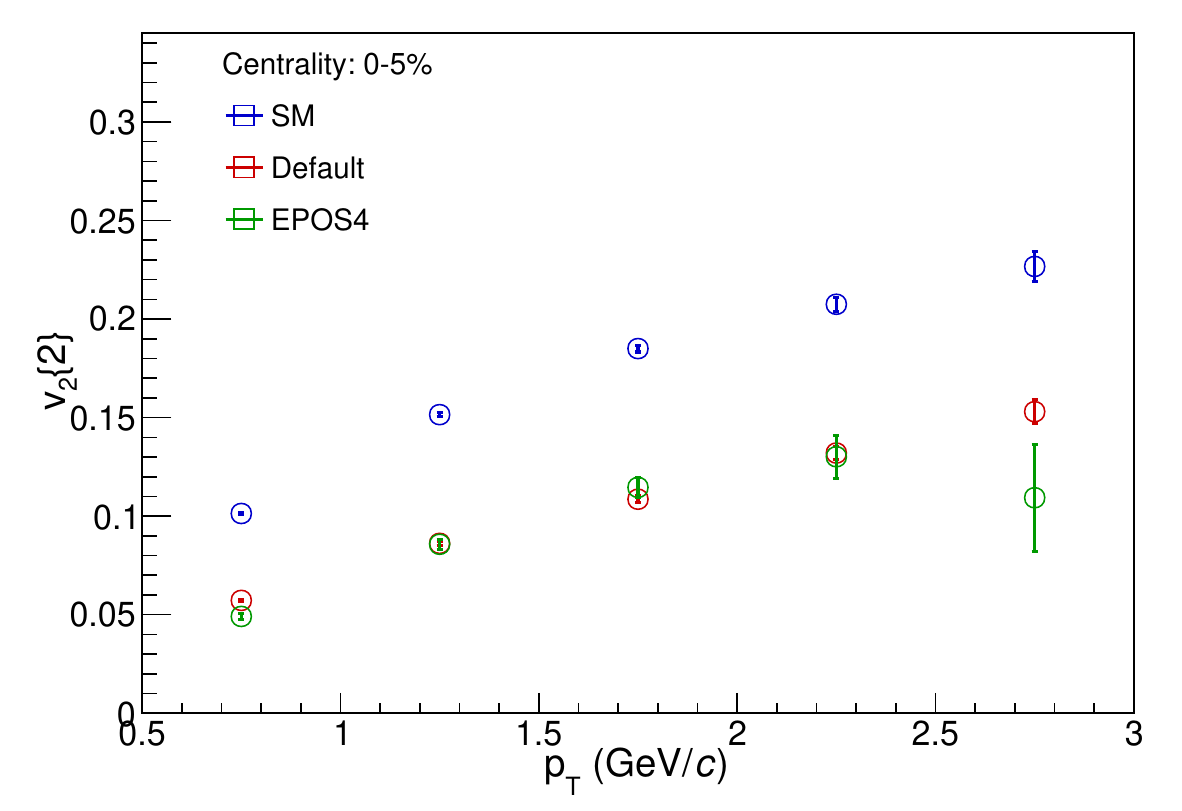}
  \includegraphics[width=.49\textwidth]{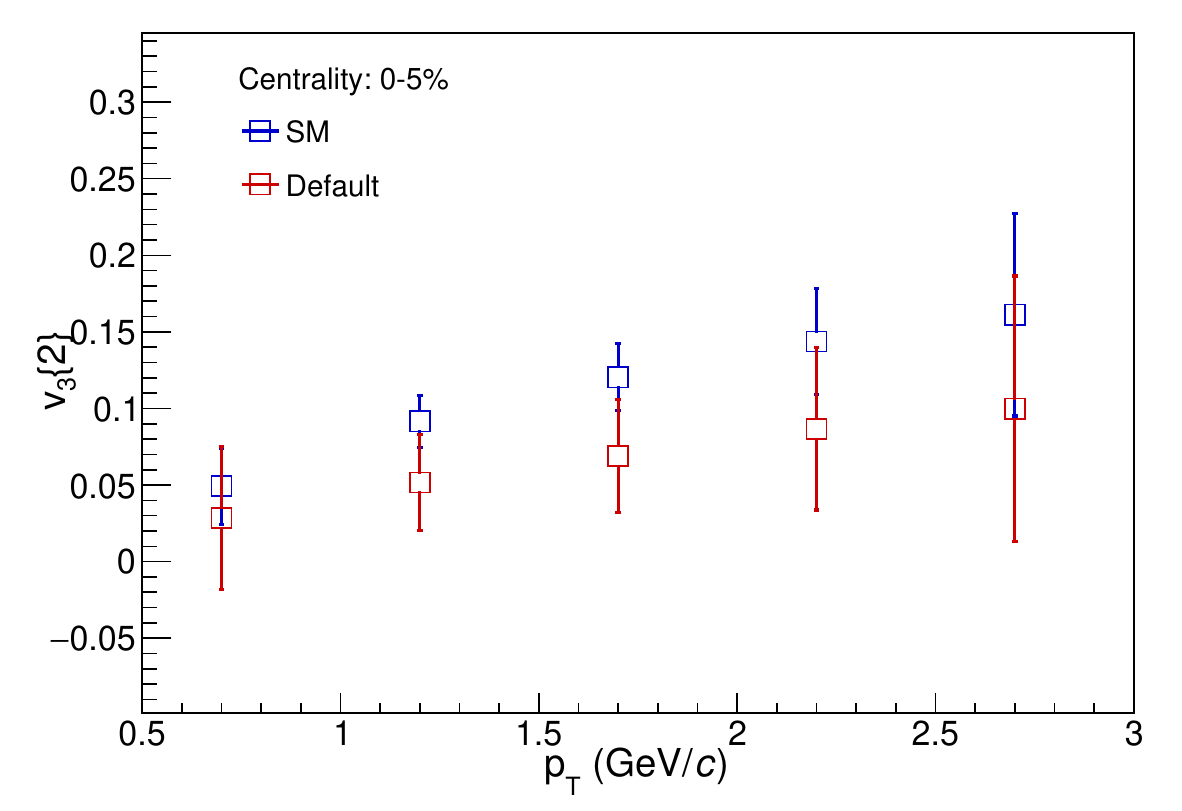}
  \caption{(Color online) $v_2$ and $v_3$ for all charged hadrons versus transverse momentum ($p_T$) in $O+O$ collisions at $\sqrt{s_{\mathrm{NN}}} = 7$ TeV, shown for 0-5$\%$ centrality using EPOS4, AMPT-Def, and AMPT-SM models.}
  \vspace{-.02cm}
  \label{fig:v2v3pt}
\end{figure}

The results of $v_2$ and $v_3$ for all charged hadrons as a function of $p_T$ in 0-5$\%$ central $O+O$ collisions using EPOS4 and AMPT are shown in Fig.~\ref{fig:v2v3pt}. Both models predict an increasing trend in $v_2$ with increasing $p_T$, while $v_2$ from AMPT-SM is higher compared to EPOS4 and AMPT-Def. $v_3$ also shows an increase in magnitude with increasing $p_T$ for both versions of AMPT model. The increasing trend of $v_2$ and $v_3$ highlights the interplay between hydrodynamic evolution and partonic interactions.

\begin{figure}[htbp]
  \centering
  \vspace{-.02cm}
  \includegraphics[width=.24\textwidth]{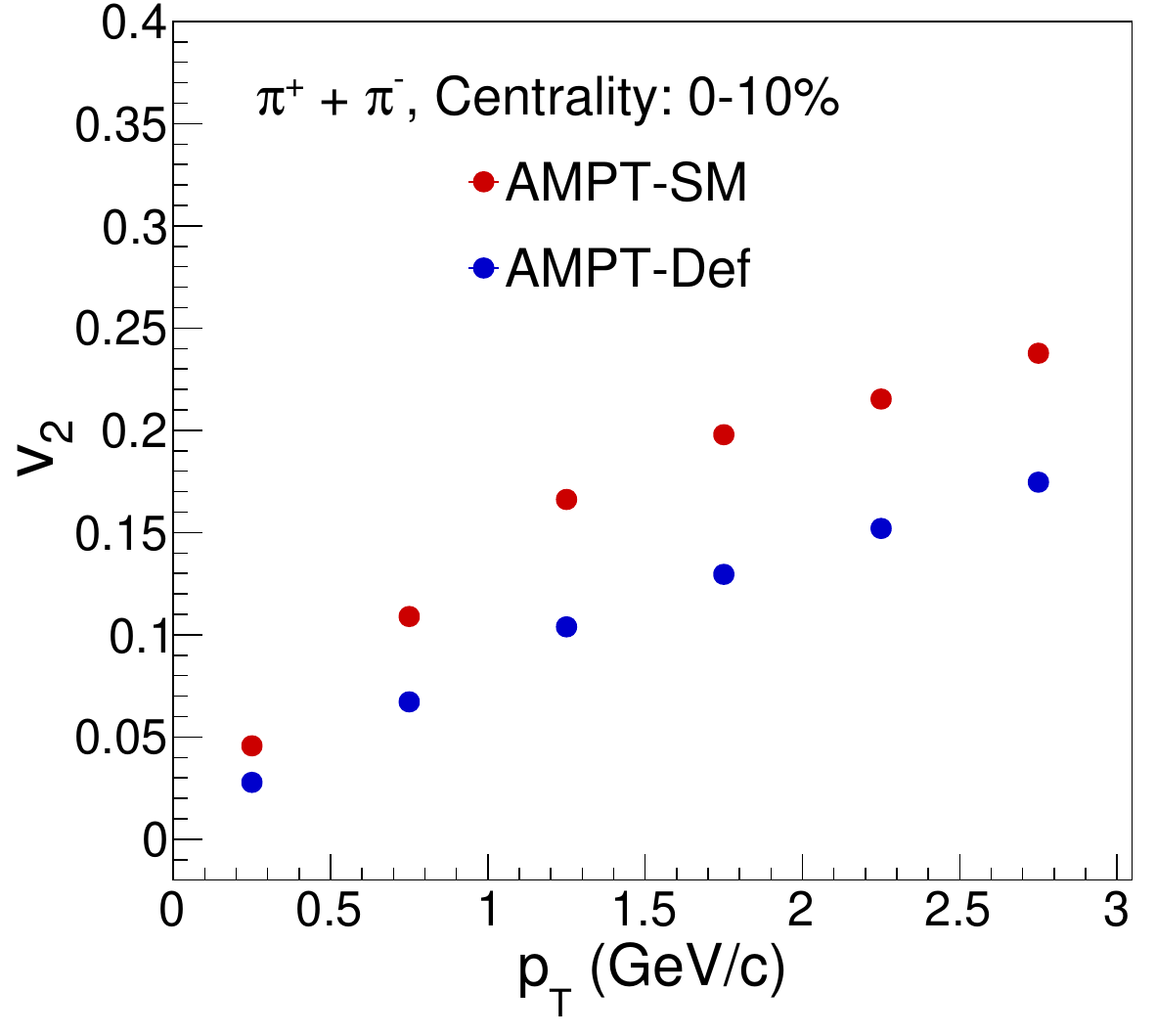}
  \includegraphics[width=.24\textwidth]{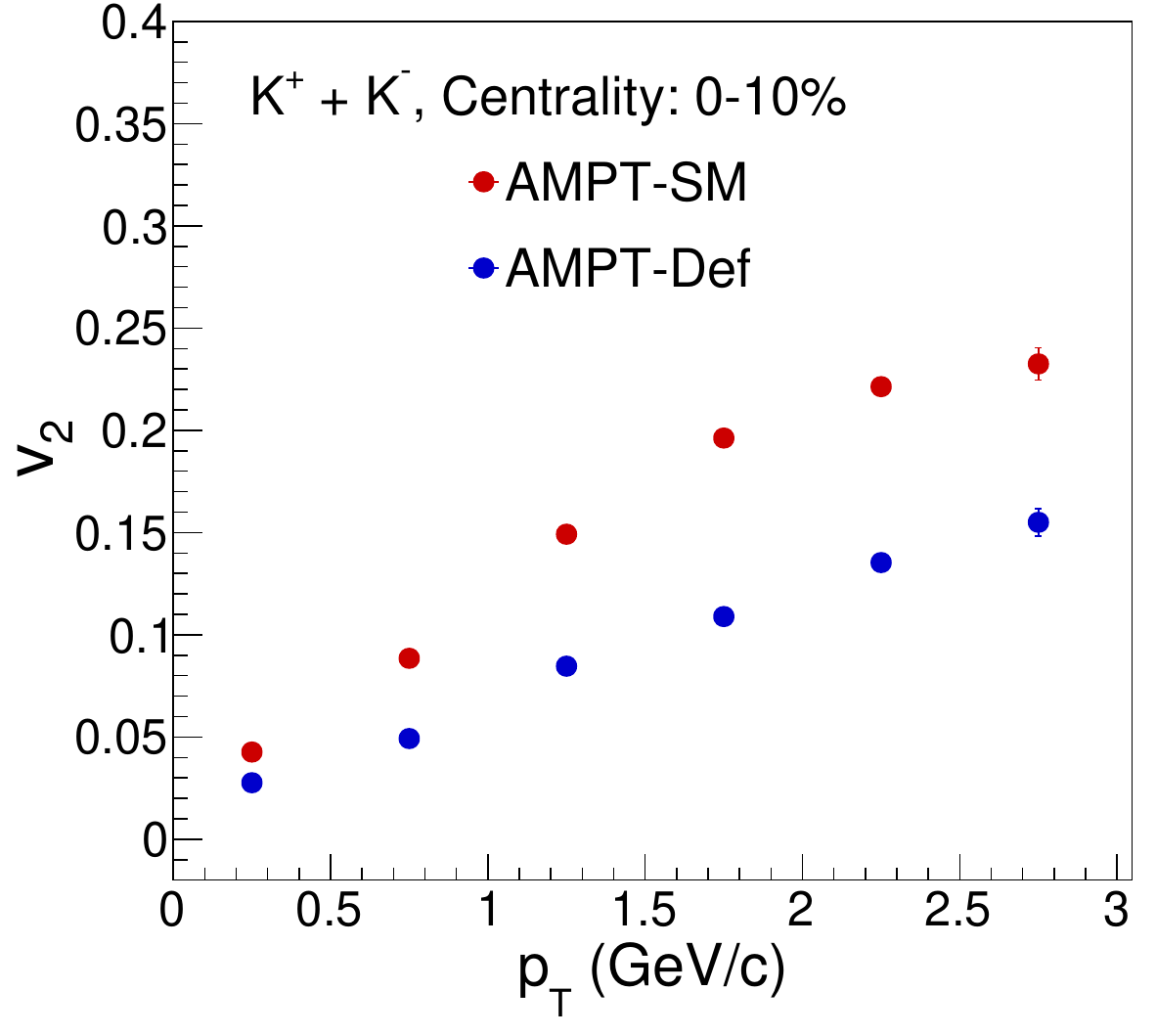}
  \includegraphics[width=.24\textwidth]{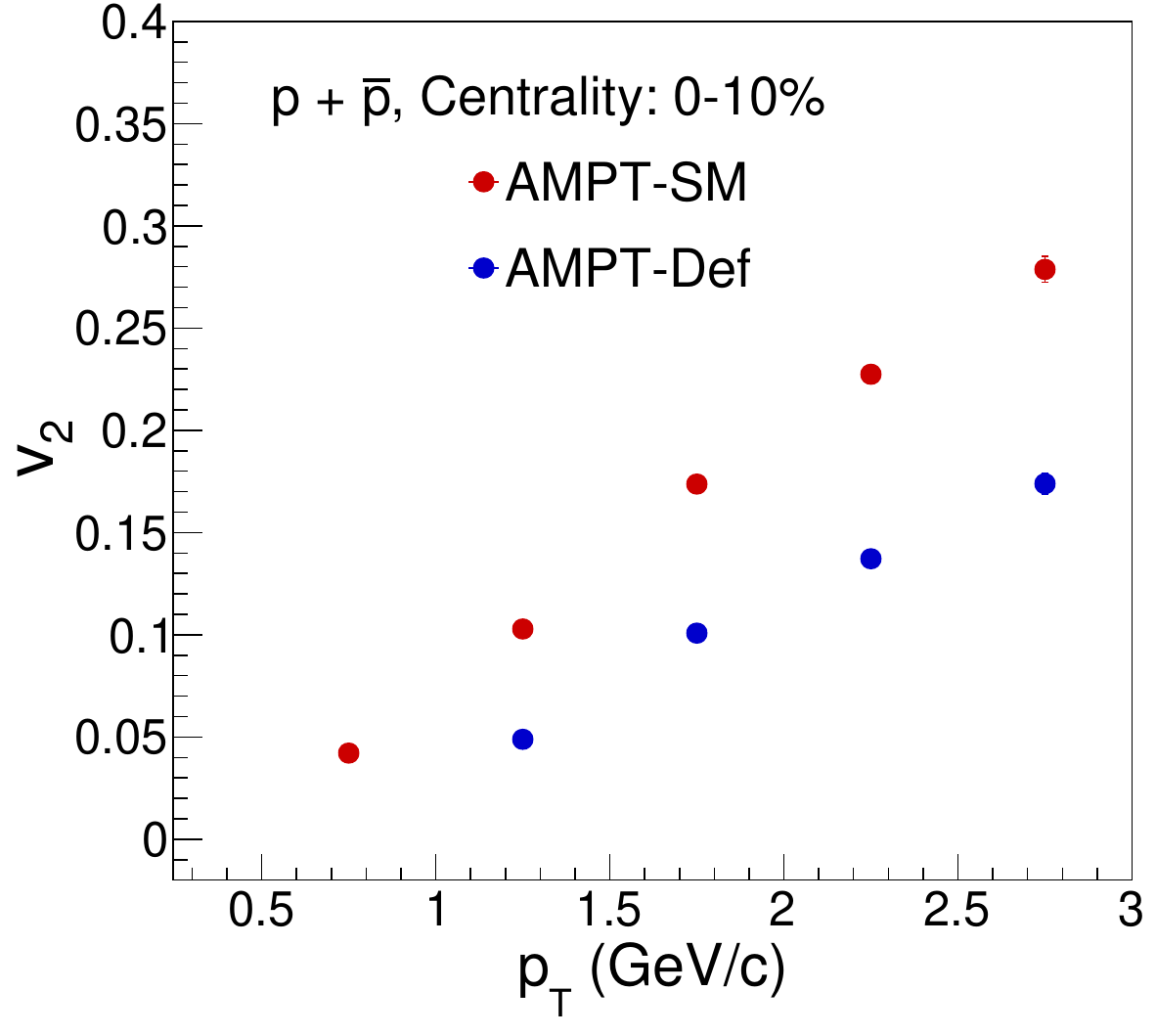}
  \includegraphics[width=.24\textwidth]{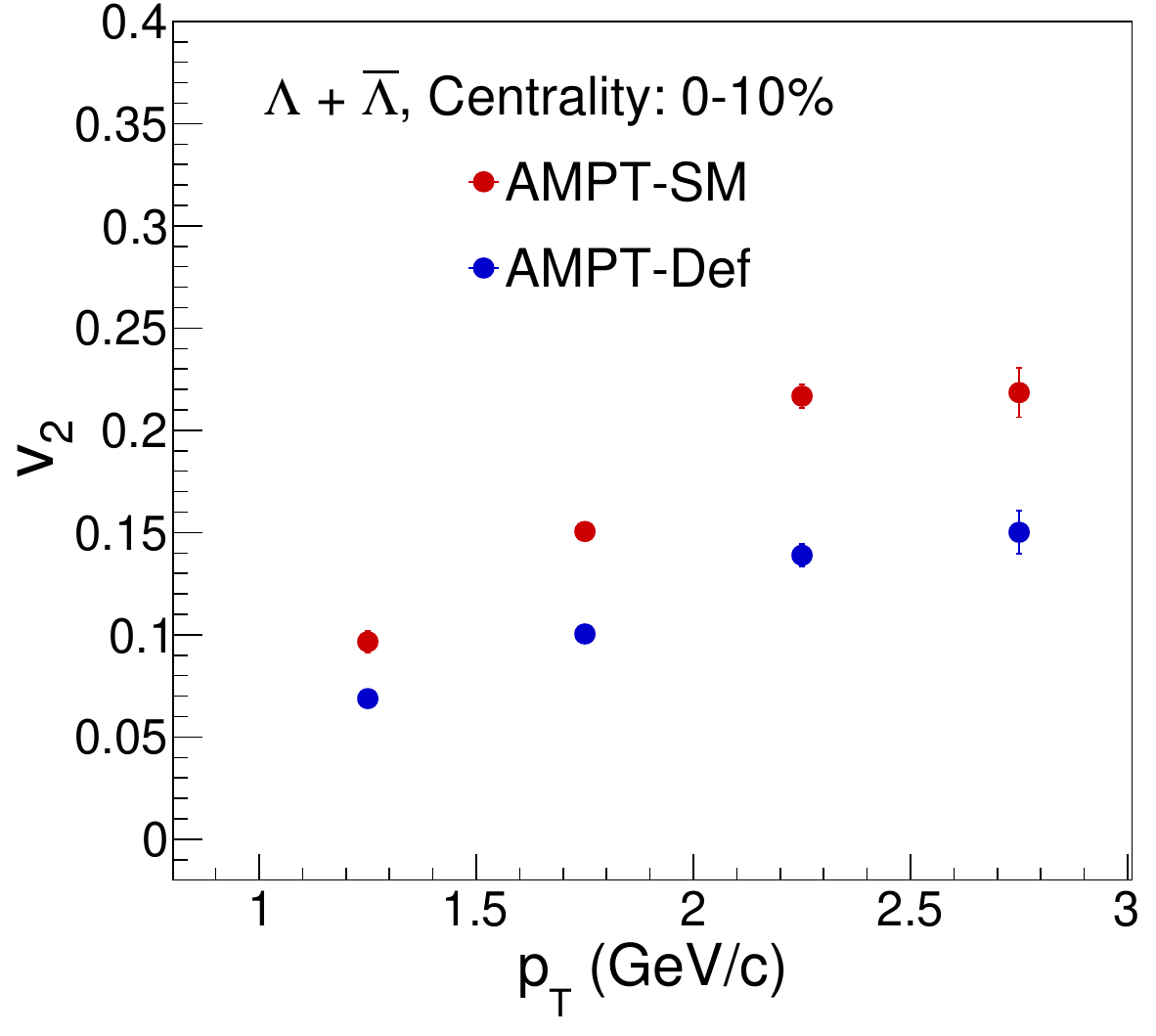}
  \caption{(Color online) $v_2$ as a function of transverse momentum ($p_T$) in $O+O$ collisions at $\sqrt{s_{\mathrm{NN}}} = 7$ TeV, shown for identified ($\pi^\pm$, $K^\pm$, $p (\overline{p})$), and strange ($\Lambda (\overline{\Lambda})$) hadrons in the 0-10$\%$ centrality class using AMPT model.}
  \vspace{-.02cm}
  \label{fig:particlept}
\end{figure}

Figure~\ref{fig:particlept} shows the predictions of elliptic flow ($v_2$) for $\pi^\pm$, $K^\pm$, $p (\overline{p})$, and $\Lambda (\overline{\Lambda})$ in  0-10$\%$ central $O+O$ collisions at $\sqrt{s_{\mathrm{NN}}} = 7$ TeV using AMPT-Def and AMPT-SM model. All particles show an increasing trend with $p_T$ and the value of $v_2$ is higher at intermediate $p_T$. A significant $p_T$ dependence is observed in elliptic flow for identified and strange hadrons.

\section{Summary}
We reported the predictions for $v_2$, $v_3$, and $v_4$ as a function of charged particle multiplicity ($N_{ch}$) and $p_T$ in $O+O$ collisions at $\sqrt{s_{\mathrm{NN}}} = 7$ TeV, using recently updated hydrodynamics-based EPOS4, and the transport model AMPT-Def and APMT-SM. The predictions for anisotropic flow ($v_n$, $n = 2, 3, 4$) from AMPT-Def is close to published experimental results from $p+p$, $p+Pb$ and $Pb+Pb$ collisions at different LHC energies. Interestingly, a final state multiplicity overlap is observed (for $v_n$) when compared with published results of the small and large systems. A strong $p_T$ dependence is observed in elliptic flow for identified ($\pi^\pm$, $K^\pm$, $p (\overline{p})$) and  starnge ($\Lambda (\overline{\Lambda})$) hadrons. Comparing these predictions with upcoming $O+O$ data at the LHC will help improve our understanding of initial-state effects and collective behavior across different system sizes.

%

\end{document}